\documentclass[12pt]{iopart}
\usepackage{longtable}
\usepackage[latin1]{inputenc}
\usepackage{graphicx,subfig}
\usepackage{color}
\usepackage{epsfig}
\definecolor{lightblue}{rgb}{0.7, 0.7, 1}
\definecolor{lightred}{rgb}{1, 0.8, 0.8}

\begin{document}

\bibliographystyle{unsrt}
\title{Ultracold quantum gases in triangular optical lattices}

\author{C Becker$^1$, P Soltan-Panahi$^1$, J Kronj\"ager$^2$, S D\"orscher$^1$, K Bongs$^2$ and K Sengstock$^1$}
\address{$^1$ Institut f\"ur Laserphysik, Universit\"at Hamburg, D-22761, Germany}
\address{$^2$ MUARC, School of Physics and Astronomy, University of Birmingham, Edgbaston, Birmingham B15 2TT, UK}
\ead{cbecker@physnet.uni-hamburg.de}

\begin{abstract}

Over the last years the exciting developments in the field of ultracold atoms confined in
optical lattices have led to numerous theoretical proposals devoted to the quantum simulation
of problems e.g. known from condensed matter physics.
Many of those ideas demand for experimental environments with non-cubic lattice geometries.
In this paper we report on the implementation of a versatile three-beam lattice allowing for the
generation of triangular as well as hexagonal optical lattices.
As an important step the superfluid-Mott insulator (SF-MI) quantum phase transition has been observed and investigated
in detail in this lattice geometry for the first time.
In addition to this we study the physics of spinor Bose-Einstein condensates (BEC) in the presence of the triangular
optical lattice potential, especially spin changing dynamics across the SF-MI transition.
Our results suggest that below the SF-MI phase transition, a well-established mean-field model describes the observed
data when renormalizing the spin-dependent interaction.
Interestingly this opens new perspectives for a lattice driven tuning of a spin dynamics resonance occurring through the interplay of quadratic Zeeman effect
and spin-dependent interaction.
We finally discuss further lattice configurations which can be realized with our setup.

\end{abstract}


\maketitle

\section{Introduction}


The physics of quantum degenerate gases in optical lattices has rapidly grown to one of the most accounted fields of atomic physics
over the last couple of years \cite{Bloch2007a}.
This new experimental environment opens the way to study many-body physics from so called mean-field dominated physics to strongly correlated regimes and
allows for a vast amount of interesting experiments ranging from Josephson junctions \cite{Albiez2005a},
quantum registers \cite{Sorensen2001a} and ultracold chemistry \cite{Ospelkaus2006a} to the implementation of
solid state models \cite{Lewenstein2007a} and experiments concerning spin oscillations in deep optical lattices \cite{Widera2005a}
to name only a few examples.
However, up to date nearly all experiments with ultracold atoms have been performed in simple cubic lattices
owing to the ease of their experimental implementation as compared to other lattice geometries \cite{Petsas1994a}.
It is only recently that the experimental and theoretical interest in more complex lattice configurations such as arrays
of fully controllable double wells \cite{Trotzky2008a,Cheinet2008a} or artificially generated gauge potentials \cite{Lim2008a,Lin2009a} appears.
In particular it has been proposed \cite{Melko2005a,Santos2004a,Wessel2005a,Wu2006a,Mathey2007a} that cold atoms loaded into triangular
or hexagonal optical lattices exhibit very rich new phases showing e.g. supersolidity \cite{Wessel2005a,Melko2005a},
quantum stripe ordered states \cite{Wu2006a}, exotic superconducting states \cite{Mathey2007a} or graphene-like physics \cite{Juzeliunas2008a,Haddad2009a}.
In solid state physics frustrated quantum magnets on triangular or trimerized lattices attract the highest amount
of interest due to their intrinsic link to high temperature superconductivity (see e.g. \cite{Nakatsuji2005a}
and references therein).
Recently it was proposed how frustrated quantum magnets can be realized within a triangular optical lattice \cite{Eckard2009a}.\\
In the context of magnetic interaction in ultracold quantum gases, spin mixtures are currently receiving rapidly growing experimental and theoretical interest.
They combine the unprecedented control achieved in single component Bose-Einstein condensates (BEC)
with intrinsic degrees of freedom. The spin-dependent coupling connects quantum gas physics to fundamental magnetic phenomena.
Earlier work analyzed the ground states and dynamical evolution of $^{23}$Na \cite{Miesner1999a,Stamper-Kurn1999a} and $^{87}$Rb
\cite{Erhard2004b,Schmaljohann2004a,Kronjaeger2005a,Kronjaeger2006a,Chang2004a,Chang2005a,Zhang2005a} spinor condensates in various ground state hyperfine manifolds.
Fascinating phenomena like spontaneous symmetry breaking \cite{Sadler2006a}, dynamical instabilities leading to pattern formation \cite{Kronjaeger2009a,Zhang2005b}
as well as dipolar effects \cite{Vengalattore2008a} have been observed in harmonically trapped spinor condensates.
However experiments dedicated to the investigation of multicomponent systems in optical lattices have been restricted to isolated 2- and 3-body spin changing
oscillations deep in the Mott-insulating regime which are inherently different from the mean-field driven dynamical evolution in harmonically trapped BEC \cite{Widera2005a,Gerbier2006a}.
The physics of spinor condensates in the presence of periodic potentials but still in the superfluid regime is thus far widely unexplored.
In this manuscript we report on a new experimental setup realizing a $^{87}$Rb BEC in a triangular optical lattice \cite{Petsas1994a}.\\
We focus on a concise understanding of the Mott-insulator (MI) transition in the triangular lattice which is a necessary precondition for further
experiments on quantum magnetism in weakly as well as in strongly correlated regimes.
In particular we work in a 2D system, where the higher dimensional degrees of freedom are frozen out by a sufficiently tight transverse confinement mediated by a 1D lattice.\\
Furthermore we investigate spin dynamics of $^{87}$Rb $F=1$ spinor BEC in this new lattice geometry.

\section{The three-beam lattice}
\begin{figure}[htb]
  \centering
  \includegraphics[width=1\textwidth]{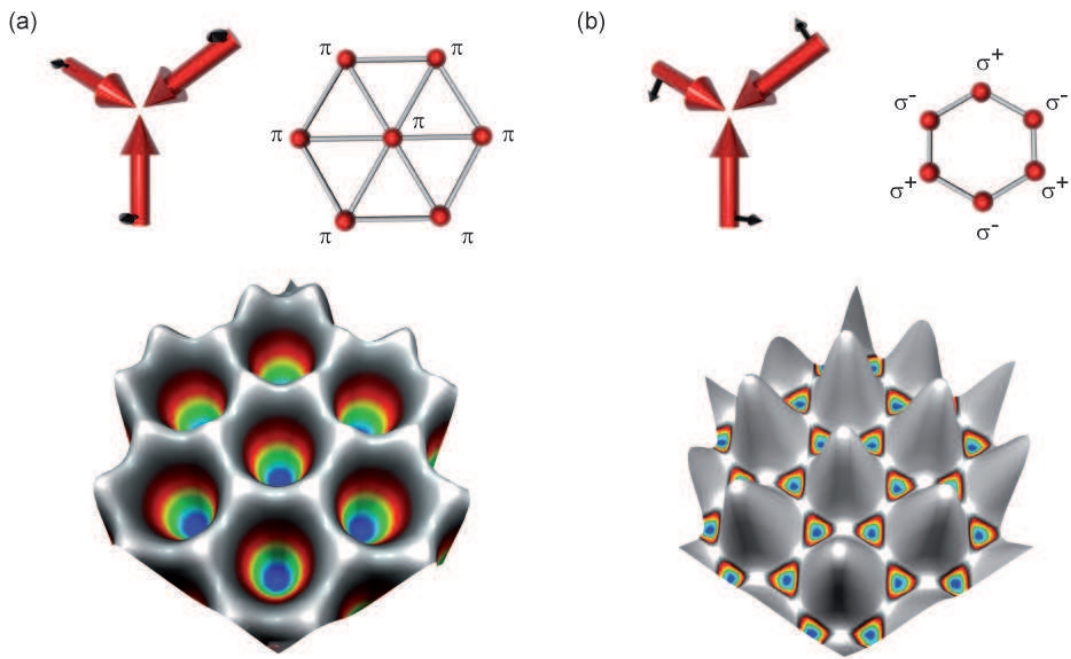}\\
  \caption[Figure1]{
    Three beams intersecting in a plane and enclosing angles of $120°$ pairwise create a triangular optical lattice
    if their polarization is perpendicular to the plane (a).
    When the polarization is rotated into the plane spanned by the lattice beams (b) a hexagonal lattice with alternating circular polarization
    at the potential minima is obtained.
    A perpendicular 1D lattice provides an overall three-dimensional periodic confinement.
    }
 \label{fig:1}
\end{figure}

\begin{figure}[htb]
  \centering
  \includegraphics[width=1\textwidth]{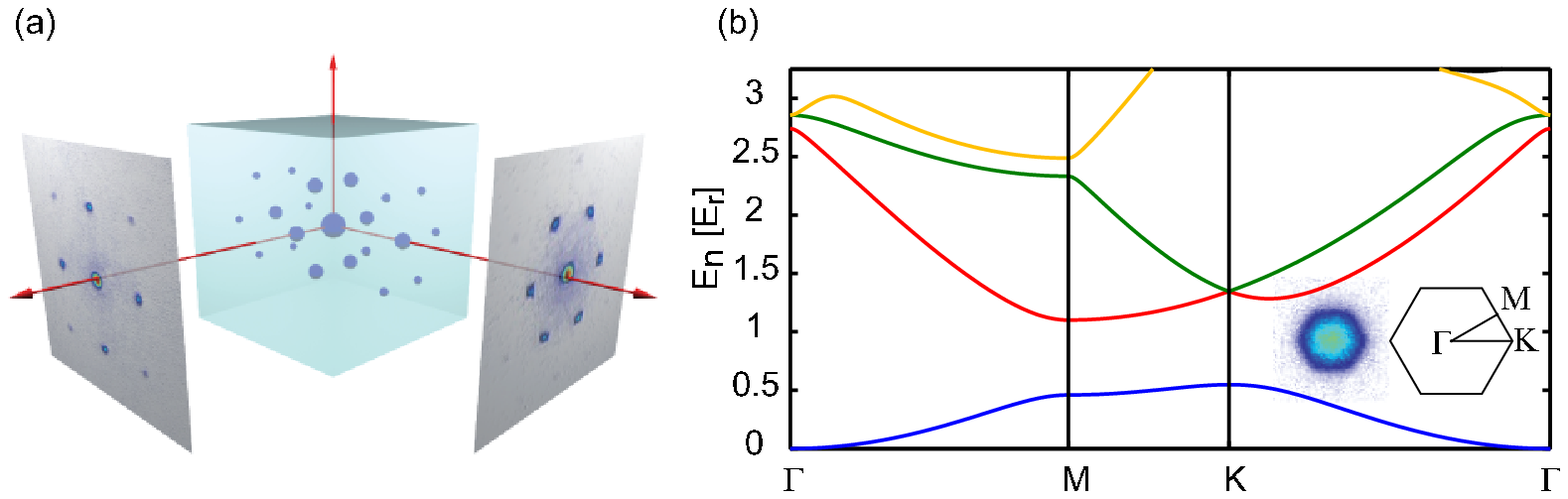}\\
  \caption[Figure2]{
    (a) Absorption images of ultracold atoms released from a triangular optical lattice.
        The sample has been imaged from two different directions
    (b) Result of a full 2D band structure calculation for the triangular optical lattice for a lattice depth of $3\,E_{\mathrm{R}}$.
        The lowest four bands are presented along lines connecting points of high symmetry as depicted in the inset.
        An experimental time-of-flight image of the Brillouin zone is also shown.
        }
  \label{fig:2}
\end{figure}

We start with a description of the experimental realization of the triangular optical lattice.
The periodic potential employed in our experiment is created by three laser beams which intersect in the $x-y$ plane mutually enclosing angles
of $120°$ with lattice vectors $\mathbf{k}_{1} = k\,(1,0)$, $\mathbf{k}_{2} = k\,(-\sqrt{3}/2,-1/2)$ and $\mathbf{k}_{3} = k\,(\sqrt{3}/2,-1/2)$ (Fig.\ref{fig:1}a).
$k$ is the wave vector of the laser employed for the lattice beams.
Adding up the field strengths of the individual beams coherently
\begin{equation}
\mathbf{E}_{\mathrm{2D}}(\mathbf{r},t) = \sum_{i=1}^{3} E_{0\,i}\ \mathbf{\varepsilon}_{i} \cos(\mathbf{k}_{i} \mathbf{\cdot r}-\omega t + \phi_i)
\end{equation}
and averaging over time results in a potential that can be written as
\begin{equation}
V(\mathbf{r}) = V_{0} \left (  \frac{3}{4} + \frac{1}{2} \left( \cos{(\mathbf{b}_1 \mathbf{r})} + \cos{(\mathbf{b}_2  \mathbf{r} + \phi_{12})} + \cos{((\mathbf{b}_1 - \mathbf{b}_2) \mathbf{r} + \phi_{23})} \right) \right)
\end{equation}
for the case that all $\varepsilon_i$'s  are the same i.e.\~ the polarization is perpendicular to the plane spanned by the lattice
beams.
For simplicity the relative phases $\phi_{ij}$ will be set to zero in the following without loss of generality.
$V_{0} = 4 I_{0}$ corresponds to the depth of a 1D lattice created by two counterpropagating  beams, where $I_{0}$ is the light shift produced
by one of the lattice beams.
The corresponding reciprocal lattice vectors $\mathbf{b}_i$ can simply be derived through the
relation $\mathbf{b}_i = \epsilon_{ijk}(\mathbf{k}_j-\mathbf{k}_k)$, while the primitive direct lattice vectors follow according
to $\mathbf{a}_i \cdot \mathbf{b}_j = 2\pi\,\delta_{i,j}$.
The resulting potential is plotted in Fig.\ref{fig:1}a and obeys a triangular symmetry.
It is important to note, that in this configuration the polarization is purely linear and the potential therefore almost $m_{F}$-independent for large
laser detuning.
A variation of the phases $\phi_i$ of the laser beams only affects the lattice through a global shift in position.
This very fact can be readily used to move, accelerate or even rotate the lattice \cite{Eckard2009a}.
In order to directly illustrate the effect of the triangular lattice structure at this point, Fig. \ref{fig:2}a shows absorption images
which have been obtained after switching off all optical potentials abruptly which projects the quasi momentum distribution in the lattice on free momentum states.
Neglecting interaction effects the density distribution after a sufficiently long time-of-flight
(TOF) therefore represents the quasi momentum distribution in the lattice \cite{Gerbier2008a}
which directly reveals the reciprocal lattice of the triangular lattice is shown in Fig.\ref{fig:2}a.
Opposed to that Fig. \ref{fig:2}b reveals the shape of the first Brillouin zone which can be obtained by first populating the whole
first band by carefully exciting the atoms followed by an adiabatic ramp-down of the lattice potential thereby mapping the quasi momentum on free momentum states.\\
By rotating the polarization of the three lattice beams {\it in} the $x-y$ plane as depicted in Fig \ref{fig:1}b, another lattice geometry can be realized.
The potential minima are now ordered on a hexagonal lattice as shown in Fig \ref{fig:1}b.
Moreover the polarization at the potential minima is perfectly circular and alternating in helicity for nearest-neighboring sites.
For alkali atoms, such as $^{87}$Rb the dipole force can be rewritten in a very instructive form
if the laser detuning $\Delta$ is on the order of the fine-structure splitting $\Delta_{\mathrm{FS}}$ in $^{87}$Rb
\begin{equation}
U_{\mathrm{dip}}(\mathbf{r}) = \frac{3 \pi c^{2}}{2 \omega_{0}^{3}}\frac{\Gamma}{\Delta} \left(1 +  \frac{1}{3} \mathcal{P} g_{F} m_{F} \frac{\Delta_{\mathrm{FS}}}{\Delta} \right) I(\mathbf{r}),
\label{equ:pot:mag}
\end{equation}
where $\omega_0$ is the laser frequency, $m_{F}$ the magnetic sub-state of the atoms and $\mathcal{P} = 0,\pm 1$ characterizes the laser polarization $(\pi, \sigma^{\pm})$ respectively.
This gives clear indication that for circular polarization the dipole potential will obey a spin dependence, which vanishes asymptotically for infinite detuning.
Taking this into account, the hexagonal lattice will induce an anti-ferromagnetic ordering for atoms with $m_{F} \neq 0$ .
To obtain a quantitative expression for the hexagonal potential it is convenient to decompose the electric field into its components in a circular basis $\mathbf{e}_{\sigma^{\pm}}$.
The corresponding potential for the two polarizations is thus given by
\begin{eqnarray}
V_{+} & = &  \frac{V_{0}}{8} \left( 3 + 2(\cos(\mathbf{b}_{1} \cdot \mathbf{r} - 2 \phi_{c}) - \cos(\mathbf{b}_{2} \cdot \mathbf{r} - \phi_{c}) + \cos((\mathbf{b}_{1} - \mathbf{b}_{2})  \cdot \mathbf{r} - 2 \phi_{c})\right)\\
V_{-} & = &  \frac{V_{0}}{8} \left( 3 + 2( -\cos(\mathbf{b}_{1} \cdot \mathbf{r} - \phi_{c}) + \cos(\mathbf{b}_{2} \cdot \mathbf{r} - 2\phi_{c}) + \cos((\mathbf{b}_{1} - \mathbf{b}_{2})  \cdot \mathbf{r} + 2\phi_{c})\right),
\label{equ:lattice:PotentialSigma}
\end{eqnarray}
with the characteristic phase $\phi_{c}=\pi/3$.\\
These two expressions clearly emphasize, that the translational symmetry of the underlying sub-lattices
is exactly the same as for the triangular lattice.
The change in polarization only corresponds to a change of the local basis with respect to the position of the individual
potential wells inside a primitive cell.\\
By finding the maxima of the above expressions one can easily show that the primitive cell of the hexagonal lattice contains an anti-ferromagnetic basis with one $\sigma^{+}$-well at
$\mathbf{a}^{\star}_1 = 1/3 \,(\mathbf{a}_1 + 2\mathbf{a}_2) $ and a $\sigma^{-}$-well at $\mathbf{a}^{\star}_2 = 1/3\,(2\mathbf{a}_1 + \mathbf{a}_2)$.
Disregarding the $m_{F}$-dependence as justified for $| F,0 \rangle$, the pure intensity modulation given by the sum of the two terms in \ref{equ:lattice:PotentialSigma} can be written as
\begin{equation}
\label{equ:lattice:hexagonal:total}
V_{\mathrm{2D}}^{\pm}(\mathbf{r}) = V_{0} \left (  \frac{3}{4} - \frac{1}{4} \left( \cos{(\mathbf{b}_1 \cdot \mathbf{r})} + \cos{(\mathbf{b}_2 \cdot  \mathbf{r})} + \cos{((\mathbf{b}_1 - \mathbf{b}_2) \cdot \mathbf{r})} \right) \right).
\end{equation}
For other $m_{F}$-states of $^{87} $Rb in $F=2$ and a laser wavelength of $830 \,\mathrm{nm}$ the relative difference in the optical dipole potential
can be as large as $(U_{\mathrm{dip}}^{|+2\rangle} -U_{\mathrm{dip}}^{|-2\rangle})/U_{\mathrm{dip}}^{|0\rangle} \approx 20\%$ according to Equ. \ref{equ:pot:mag}.
We emphasize that the interplay between spin-dependent contact interaction, frustrated inter-site dipolar interaction and a spin-dependent
trapping potential where e.g. spin-dependent tunneling could be realized may give rise to a whole range of new physical effects and exotic ground state phases.\\

\section{Experimental preparation of BEC and lattice calibration}

In the following we briefly summarize the experimental sequence employed for the measurements presented here.
BEC is produced by first collecting up to $10^{10}$ $^{87}$Rb atoms in a double magneto-optical trap system, which are subsequently loaded in a
magnetic trap of hybrid D-cloverleaf type.
Evaporative cooling leaves us with about $1.5 \times 10^{6}$ ultracold atoms slightly above the critical temperature for Bose-Einstein condensation.
Afterwards the atoms are transferred to a crossed dipole trap with typical trapping frequencies  $\bar{\omega} = 2\pi \times (90 \pm 3) \mathrm{Hz}$.
The large detuning of the employed Nd:YAG laser with a wavelength of $\lambda = 1064\,\mathrm{nm}$ ensures a spin state independent global trapping
allowing for the investigation of spinor physics at all lattice depths.
Further evaporative cooling by smoothly lowering the intensity of the dipole trap laser beams produces BEC with particle
numbers ranging from $6$ to $7 \times 10^{5}$ without any discernible thermal fraction.
Since we are especially interested in the physics induced by the triangular lattice we eliminate the influence of the third dimension
by creating a stack of quasi two-dimensional disc-shaped condensates.
This is done  by ramping up a 1D lattice over $150\,\mathrm{ms}$ to a final depth of $V_{0} = 30 E_{r}$,
where the recoil energy is given by $E_{r}= (\hbar k_{\mathrm{lattice}})/2m_{\mathrm{Rb}}$.
After this step the tunneling along the perpendicular direction $\hbar/J_{1D} \approx 0.5\,\mathrm{s^{-1}}$ is negligible compared
to the time needed for the experiment $\tau_{\mathrm{exp}} = 0.17\,\mathrm{s}$.
The condition for being in the so called quasi two-dimensional regime which requires
$k_{\mathrm{B}}T,\,\mu \ll \hbar \omega_{1D} = \hbar \cdot 2\pi \cdot 21\,\mathrm{kHz} = k_{\mathrm{B}} \cdot 1\,\mu \mathrm{K}$ is also well fulfilled.\\
The increased effective coupling $\tilde{g} \approx 2.5 g_{0}$ leads to a slight expansion of the condensate in all directions,
whereas the enhanced harmonic confinement $\tilde{\omega} = 1.2 \bar{\omega}_{0}$ tends to compress the sample.
After all we fill about 40 discs, where the occupation of the central disc amounts to $N_{2D} \approx 4000$.
The chemical potential is $\mu = h \cdot 2.7 \,\mathrm{kHz} = 133\,\mathrm{nK}$.
Subsequently the triangular 2D lattice is ramped up to the desired final lattice depth using a smooth sigmoidal ramp within
$150\,\mathrm{ms}$ \cite{Jaksch2002a}.
After a short hold time of $5\,\mathrm{ms}$ all optical potentials are switched off within less than $10\,\mu\mathrm{s}$ and the atomic cloud
is imaged on a CCD camera after $21\mathrm{ms}$ TOF.\\
The lattice laser beams are derived from a frequency and intensity stabilized Ti:Sa laser, which delivers up to $1.3\,\mathrm{W}$ of power
at a wavelength of $830\,\mathrm{nm}$.
Beam shaping telescopes focus the laser beams to Gaussian waists of $125\,\mathrm{\mu m}$.
Special care has to be devoted to the relative phases of the three laser beams generating the triangular lattice.
Since optical fibers are used to deliver the optical power to the experiment fast servo loops eliminating any phase noise imprinted by acoustic Brillouin
scattering inside the fibers are employed.
This avoids heating caused by parametric excitation \cite{Petsas1994a,Savard1997a}.\\
We calibrate the lattice depth by parametric excitation \cite{Savard1997a} from the first to the
third band at sufficiently large laser power, where the bandwidth of the bands becomes negligibly small, and comparison of
the observed frequencies with band structure calculations (Fig. \ref{fig:2}b shows a full 2D band structure calculation).
For this purpose we create 1D lattices using two of the three laser beams and permute through all pairs of beams
to create a potential as symmetric as possible.\\

\section{Observation of the SF-MI transition}


The SF-MI transition of ultracold atoms confined in an optical lattice has been under vast investigation since
first pioneering theoretical \cite{Fisher1989a,Jaksch1998a} and experimental \cite{Greiner2002a} studies over the last decade.
Many aspects and characteristic features of this quantum phase transition have been addressed such as the visibility of
interference peaks \cite{Gerbier2005a,Gerbier2005b,Zwerger2003a,Gerbier2008a},
the possibility of observing fine structure in the central peak \cite{Kashurnikov2002a,Wessel2004a},
probing of the energy gap in the MI phase \cite{Greiner2002a}
and as a striking demonstration of strong correlations in the MI phase noise correlations have been observed in various experiments
\cite{Foelling2005a}.
Moreover continuative experiments could directly map the wedding cake like shell structure of different
single-site occupation numbers in the MI by high resolution tomography \cite{Foelling2006a}, high accuracy rf-spectroscopy
\cite{Campbell2006a} and direct in-situ imaging \cite{Gemelke2009a}.
The effects of finite temperature \cite{Gerbier2007a,Pupillo2006a,Kato2008a,Yi2007a} and dimensionality
\cite{Stoeferle2004a} have also been in the focus of interest.
A detailed understanding of the MI state is an indispensable requirement for many fascinating experiments that have
been proposed in the field of quantum information processing \cite{Calarco2004a} and simulation of solid state systems
\cite{Lewenstein2007a}.

\begin{figure}[htb]
  \centering
  \includegraphics[width=1\textwidth]{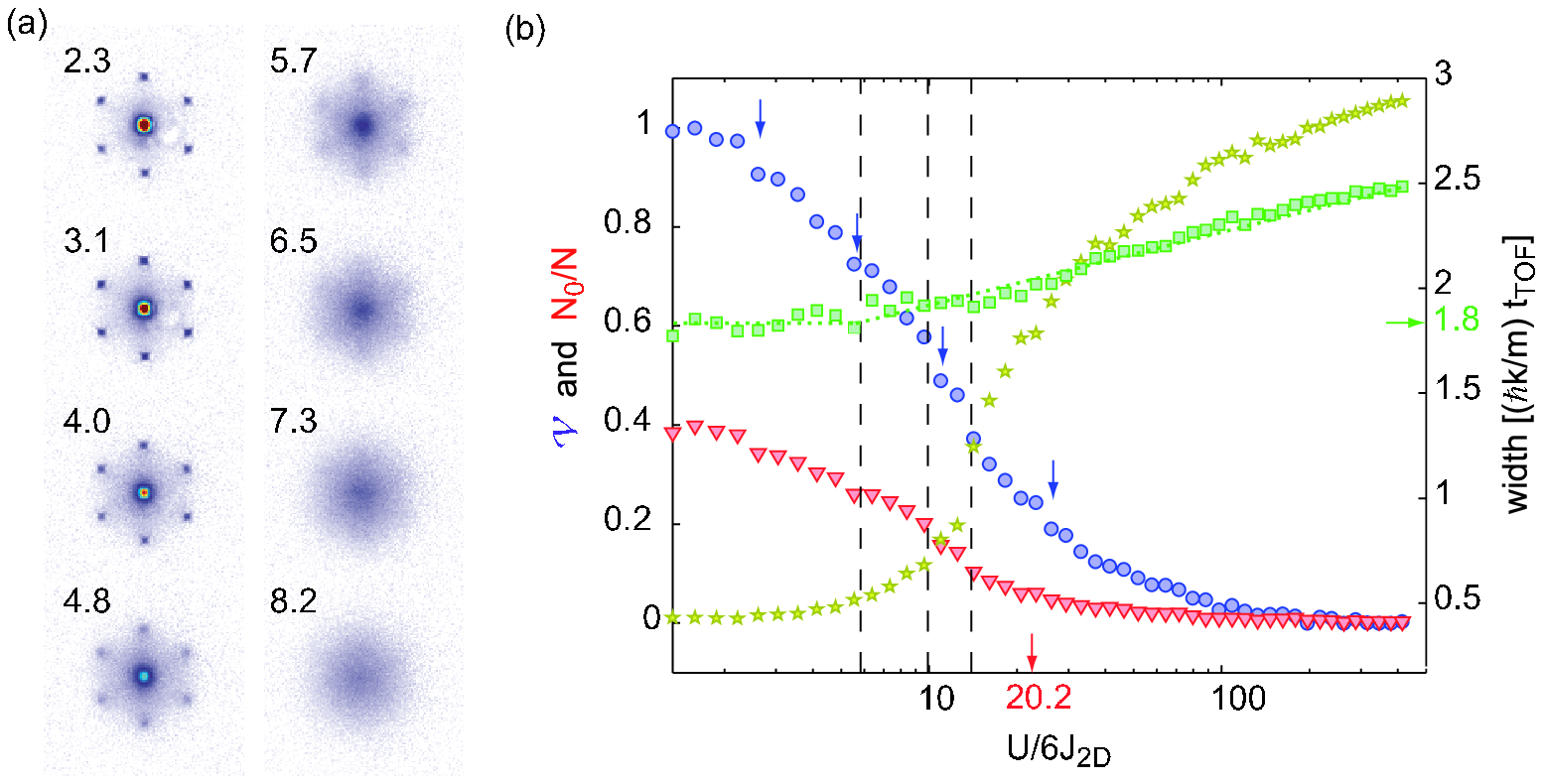}\\
  \caption[Figure3]{
    (a) Absorption images of atoms released from a triangular optical lattice for various lattice depths given in units of $E_{r}$.
    (b) Visibility of the interference peaks (blue circles) across the phase transition against $U/zJ$.
        Note that the visibility has been normalized to its maximum value of $\mathcal{V}_{\mathrm{max}} = 0.48$.
        The visibility starts to drop around a lattice depth of $3.5 -4 \,\mathrm{E_{r}}$.
        Reproducible kinks marked by blue arrows may indicate the formation of Mott-shells.
        The condensate fraction (red triangles) decreases as the lattice is ramped to deeper values.
        Pronounced kinks coinciding with those observed for the visibility are observed.
        The condensate fraction vanishes more quickly than the visibility ($N_{0}/N = 0$ indicated by red arrow).
        The FWHM of the central peak of the momentum distribution (yellow stars) is constant for low lattice depth.
        Around the phase transition it quickly starts to grow -- also exhibiting small but reproducible kinks.
        The width of the incoherent background is plotted for completeness (green squares).
        Constant below the phase transition where it is a measure for the temperature (indicated by a green arrow),
        it starts to increase linearly with respect to $U/zJ$ in the MI regime as expected from perturbation theory.
        Dashed vertical lines indicate the critical values for the transition to a Mott insulating state with $\bar{n}=1,2$ and $3$
        respectively.
            }
  \label{fig:3}
\end{figure}

Following the experimental sequence outlined above we have investigated the SF-MI transition in a triangular optical
lattice for the first time.
Fig \ref{fig:3}a shows a series of time-of-flight absorption images obtained by varying the lattice depth across the SF-MI transition.
We have checked the reversibility of the ramp process by first bringing the system deep in the MI regime and
subsequently ramping the lattice down to a depth corresponding to maximal visibility and holding the atoms for
a short time comparable to the tunneling time $J/\hbar$, after which we observe a very high recurrence of the visibility.

\subsection{Visibility}

The visibility, determined as described in \cite{Gerbier2005a} and averaged over $15 - 20$ experimental realizations
is shown in a (semi-logarithmic) plot versus the Hubbard parameter $\eta$ in Fig. \ref{fig:3}b.
The visibility of the diffraction peaks corresponding to reciprocal lattice vectors start to smear out at a lattice depth
of about $3.5\,\mathrm{E_R}$ corresponding to a Hubbard parameter $\eta = U/zJ$ close to the expected value of $\eta = 5.8$ \cite{vanOosten2001a} and drops to very small values
for lattice depth $V_{0} \ge 8\,E_{\mathrm{r}}$.
Note that the number of nearest neighbours $z$ enters the calculation and is not consistent with any other number than $z=6$, which is expected for the triangular lattice.
In an inhomogeneous system a typical feature of a perfect atomic Mott-insulator is a wedding cake like structure of
the occupation number per lattice site \cite{Jaksch1998a}.
The relatively strong inhomogeneity in our experiment resulting from the large trapping frequencies of the dipole trap,
suggests a rather pronounced structure of alternating MI and superfluid shells
\cite{Gerbier2005a,Gerbier2005b,Foelling2006a,Campbell2006a,Mitra2008a} with a central site occupation $\hat{n}$ of $3-4$ atoms.
Consequently reproducible kinks in the visibility are observed as also reported in \cite{Gerbier2005a}.
The locations of the kinks fit well to the critical values of the Hubbard parameter $\eta_{\mathrm{c}}^{n}$
for the transition to a MI with occupation number $n$ derived from a mean-field model \cite{vanOosten2001a}.\\

\subsection{Condensate fraction}

As the visibility starts to drop the central as well as the diffraction peaks begin to broaden significantly as can also be deduced from Fig\ref{fig:3}b.
We have fitted a bimodal distribution with a fixed condensate width
\footnote{The condensate width has been determined using fits to data corresponding to low lattice depths well below the phase transition.}
to the peripheral peaks of the interference pattern.
In order to account for interaction induced broadening, data from the first Brillouin zone has been excluded \cite{Yi2007a,Spielman2008a}.
After sufficiently long time-of-flight the density distribution consists of sharp interference peaks \cite{Gerbier2008a} caused
by the condensed atoms which are assumed to exhibit the same parabolic shape as the external harmonic confinement.
The other part of the bimodal density distribution is produced by thermal atoms whose distribution has been modeled employing a Gaussian.
For small lattice depths this seems reasonable for a bimodal BEC in a harmonic trap \emph{perturbed} by a periodic potential.
For deep lattices the envelope of the density distribution  is given by the Fourier transform of the Wannier function $|w(\mathbf{q})|^2$
which is again well approximated by a Gaussian.
\begin{figure}[htb]
  \centering
  \includegraphics[width=1\textwidth]{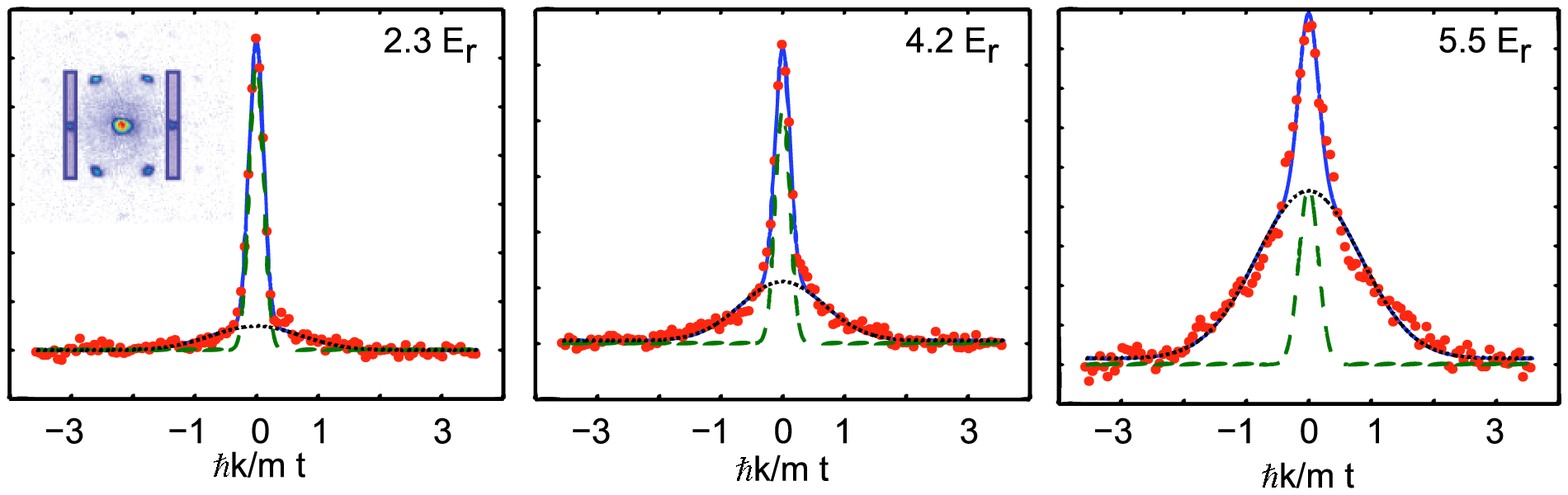}\\
  \caption[Figure5]{
		Bimodal fits to the peak structure of the absorption images.
		Data from the first Brillouin zone has been excluded to minimize interaction effects during the expansion.
		The data points consequently represent an average of all 6 satellite peaks (blue shaded area in inset).
		Note that every absorption picture has been averaged over 20-30 experimental runs.
		From left to right the lattice depth increases, reaching the critical point for singly occupied lattice sites
		in the last picture.
    }
  \label{fig:5}
\end{figure}
The condensate fraction determined in this way is also plotted and accurately follows the decrease of the visibility in shape for shallow
to intermediate lattices.
This is expected since the contrast of the interference pattern is intrinsically connected to the superfluid density \cite{Roth2003a}
and therefore the condensate fraction.

\subsection{Residual visibility - Particle-hole excitations}

\begin{figure}[htb]
  \centering
  \includegraphics[width=1\textwidth]{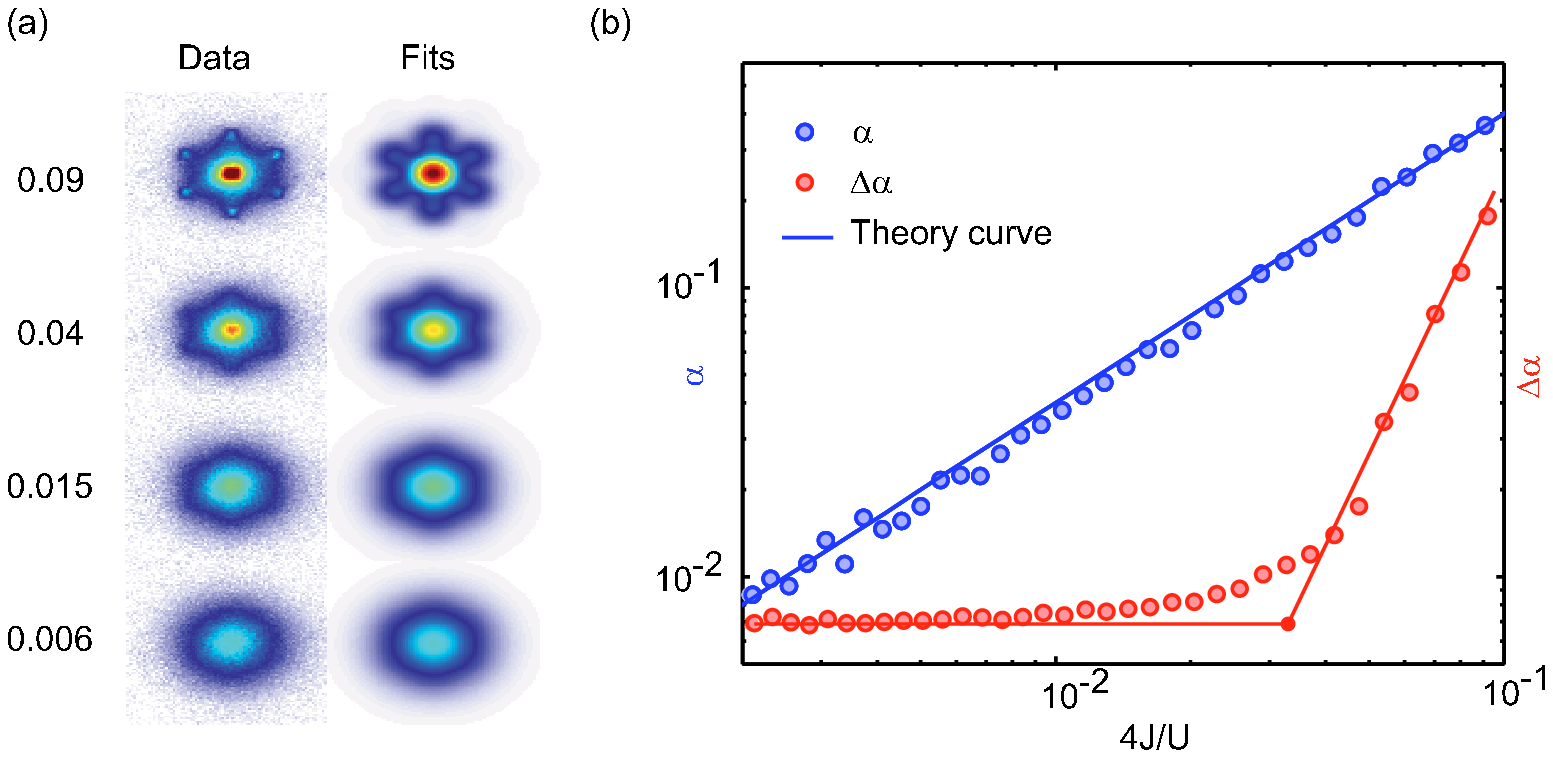}\\
  \caption[Figure4]{
    (a) Absorption images of ultracold atoms released from a triangular optical lattice for selected lattice depth above the MI transition point
        are shown together with fits of first order perturbation theory with only one free parameter $\alpha$.
    (b) Experimentally obtained values for $\alpha$ together with the theoretical expected curve for an average occupation of $\langle \hat{n} \rangle = 2$.
        Note that $\alpha$ is proportional to $8J/U$.
        The fit error $\Delta \alpha$ exhibits a distinct increase at a Hubbard parameter corresponding to the appearance of a finite superfluid
        fraction as indicated by the red arrow in Fig.\,\ref{fig:3}.
        The red solid lines are guides to the eye.
        }
  \label{fig:4}
\end{figure}

For lattice depths close to the expected SF-MI phase transition the superfluid fraction disappears while a small residual visibility persists even far in the MI regime.
This residual visibility in the MI regime has been attributed to excitations in terms of particle-hole pairs \cite{Gerbier2005a,Gerbier2005b} which should in principle
be present at any finite hopping $J$.
We have determined the coefficients $\alpha$ by fitting the model density derived from a first order perturbed MI state \cite{Gerbier2005a,Spielman2007a} according to
$\langle \hat{n}_{\mathbf{k}}\rangle = N |w(\mathbf{k})|^2\,(1+\sum_{i}\alpha \cos(\pi \mathbf{k}/\mathbf{k}_{i}))$.
The expected value $\alpha = 4J/U \hat{n}(\hat{n}+1)$ fits surprisingly well the observed data for an average
occupation number of  $\hat{n}=2$ in spite of the rich shell structure that is expected in our experiment with occupation numbers
ranging from $1$ to $4$ (see Fig.\ref{fig:4}).
The error of the applied fit starts to increase significantly (indicated by the intersection of the two red lines in Fig. \ref{fig:4}\,b)
at a lattice depth corresponding to the appearance of a finite superfluid fraction
supporting the idea of particle-hole pairs as the origin of the residual visibility.\\
Although recent simulations \cite{Kato2008a} show that ultracold thermal atoms also exhibit a significant interference pattern,
our experimental parameters suggest that loading the atoms into intermediate deep lattices should even result in adiabatic cooling \cite{Blakie2004a}.
However, the vanishing of the interference pattern has to be evaluated with great care for atomic samples with a non-negligible thermal fraction
(see \cite{Blakie2004a,DeMarco2005a,Ho2007a,Gerbier2008a}).
By ramping back from a deep MI state to zero lattice depth and analysis of the density distribution after TOF we have checked
that only negligible irreversible heating occurred for short to intermediate hold times.\\

Note that the broadening of the diffraction peaks as well as the condensate fraction also show a kink like behavior supporting
the observation derived from the visibility measurement \cite{Wessel2004a}.

\section{Spin dynamics in a triangular lattice}

We have performed measurements of spin dynamics in lattices of different depth and at different magnetic fields
in order to gain insight into the physics of magnetic systems in periodic potentials.
In principle one would expect a crossover from mean-field regime spin dynamics as described in e.g. \cite{Kronjaeger2005a} to few particle
spin oscillations isolated at individual lattice sites \cite{Widera2005a}.
It has been found for the latter case that coherent oscillations between two two-particle states $\psi_{i} = |m_{F}=0,0 \rangle \leftrightarrow \psi_{f} = |-1,1 \rangle$
occur that can be readily explained within the framework of a Rabi-like model \cite{Widera2005a}.
For atoms with $F=1$ in an optical lattice the effective Rabi frequency is composed out of a coupling strength
$\Omega_{if} \propto g_{1}\int|w(\mathbf{r})|^{4}\,d\mathbf{r}$ given by the spin-dependent interaction between the two states and a detuning $\delta = \delta_{0} + \delta(B^2)$.
$\delta$ is given by a part proportional to the quadratic Zeeman shift $\delta(B^2)$ and an offset term determined by the difference in interaction
energy of the initial and final state $\delta_{0} \propto g_{1}/2$.\\
Here we focus on the investigation of spin dynamics at intermediate lattice depths which is associated with an overall spin \emph{independent} trapping potential
from the beginning.
We present measurements of spin dynamics of $^{87}$Rb BEC in $F=1$ starting in an almost isotropic crossed optical dipole trap and ranging across
the superfluid regime and up to the SF-MI phase transition inside a triangular optical lattice.
In addition the magnetic field has been varied in order to change the relative influence of quadratic Zeeman effect and spin dependent interaction.
We show that this ratio determines the behavior of oscillation amplitude and period.
After a short reminder on the physics and notations of mean-field spinor physics within the single mode approximation and a description of the experimental sequence,
we will present results suggesting a new and elegant method to tune the spin dynamics resonance known from experiments in harmonic traps (see e.g. \cite{Kronjaeger2005a,Chang2005a}).

\subsection{{\it Mean-field} theory of spin dynamics}

The description of spinor condensates presented here is based on the evolution due to the corresponding mean-field Hamiltonian
using the single mode approximation (SMA) as outlined in more detail in e.g. \cite{Kronjaeger2005a,Kronjaeger2006a}.
Spin mixing occurs as a consequence of two-particle collisions and can be classified by the total spin $f$ of the colliding pair.
For $F=1$ spinor condensates the allowed values are $f=0,2$.
The spin mixing interaction can be parameterized by the coupling coefficient $g_1=\frac{4\pi \hbar^2}{m}\frac{a_2-a_0}{3}$,
each total spin channel being associated with an s-wave scattering length $a_f$.
The dynamics of the $2F+1$ components of the vectorial spinor order parameter directly follows from the resulting mean-field energy
functional for the spin part of the wave function $\mathbf{\zeta}$:

\begin{equation}
\label{equ:spinor:SGPE}
  \langle H_\mathrm{spin} \rangle =\frac{\hbar}{2} N\langle n \rangle \Big(g_1\langle \mathbf{F} \rangle^2  \Big) + \hbar N\Big(- p\langle F_z \rangle + q\langle F_z^2\rangle\Big)
\end{equation}
with
\begin{eqnarray}
\langle F_z \rangle   &\equiv& \sum_{ij}\zeta_i^{*}(F_z)_{ij}\zeta_j \\
\langle F^2_z \rangle &\equiv& \sum_{ijk}\zeta_i^{*}(F_z)_{ik}(F_z)_{kj}\zeta_j
\end{eqnarray}

$|p| = \frac{1}{2}\frac{\mu_B}{\hbar}B_0$ and $|q| = \frac{p^2}{\omega_\mathrm{hfs}}$ parametrize the linear and quadratic Zeeman effect respectively.
The dynamics governed by the above Hamiltonian can be solved exactly for $F=1$ and a particular initial state $\zeta_{\pi/2}$ which can be easily prepared experimentally
by the application of an initial  $\pi/2$ rf-pulse \cite{Kronjaeger2005a,Kronjaeger2006a}.
The solution of the resulting equation of motion for the populations of the individual $m_{F}$-states can be expressed in terms of Jacobi elliptical functions (JEF)
\begin{eqnarray}
  |\zeta_0(t)|^2        & = & (1-k\,\mathrm{sn}^2_k(qt))/2,\\
  |\zeta_{\pm 1}(t)|^2  & = & (1+k\,\mathrm{sn}^2_k(qt))/4.
\label{equ:spinor:pop}
\end{eqnarray}
The occurrence of the parameter $k = g_1\langle n \rangle / q$ in the analytic solution indicates that amplitude and oscillation period
of spin dynamics are determined by the ratio of interaction to quadratic Zeeman energy.
\begin{figure}[htb]
  \centering
  \includegraphics[width=1\textwidth]{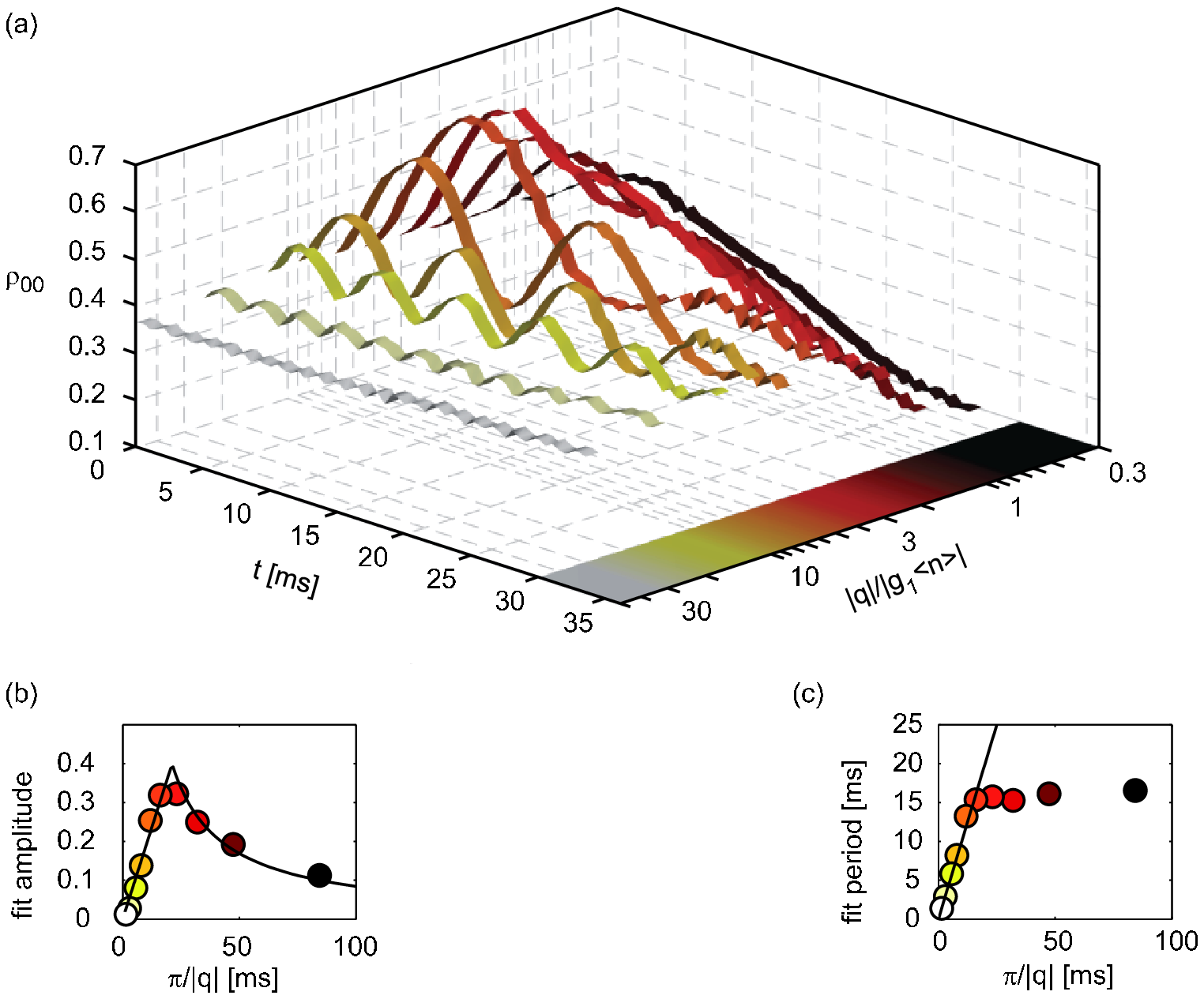}\\
  \caption[Figure6]{
  (a) Spin dynamics resonance phenomenon in $F=2$ taken from \cite{Kronjaeger2006a}.
  Plotted are coherent oscillations for different values of  $q/(g_1 \langle n\rangle)$, where $q\sim B^2$ is the quadratic Zeeman energy and
  $g_1\langle n \rangle $ the first spin-dependent mean-field energy (processes involving $\Delta m_F = 1$).
  In this figure, $g_1 \langle n \rangle = 47\,s^{-1}$ (corresponding to a density of $1 \times 10^{14}\,\mathrm{cm}^{-3}$) is used as a reference value,
  since only the magnetic field has been varied among the different data sets.
  The mean density has been obtained from SMA theory fits.\\
  Amplitude (b) and period (c) have been extracted for qualitative comparison with theoretical predictions for $F=1$.
  }
  \label{fig:6}
\end{figure}
The JEF in Equ. \ref{equ:spinor:pop} can further be simplified by approximations in two limiting cases:
\begin{description}
	\item[Zeeman regime ($k \ll 1$)]
	In the Zeeman regime, where the quadratic Zeeman effect dominates over spin-dependent interaction, the JEF can be approximated by ordinary
	trigonometric functions  $\mathrm{sn}_k(x)\approx\sin(x)$, $\mathrm{cn}_k(x)\approx\cos(x)$, $\mathrm{dn}_k(x)\approx 1$, yielding
	\begin{equation}
	\label{equ:spinor:QZR}
	  |\zeta_0(t)|^2       \,\,  \stackrel{ZMR}{=} \,\, (1-k\sin^2(qt))/2,
	\end{equation}
The amplitude of the oscillation is  proportional to $k$, while the period is simply given by $\pi/q$.
	\item[Interaction regime ($k \gg 1$)]
	In this regime the spin dependent interaction dominates any dynamics.
	The appropriate approximations for the JEF read \newline
	$\mathrm{sn}_k(x) = \frac{1}{k}\sin_{1/k}(kx)$ resulting in population oscillations
	\begin{equation}
	\label{equ:spinor:IAR}
	  |\zeta_0(t)|^2     \,\,    \stackrel{IAR}{=} \,\, (1-1/k\sin^2(g_1\langle n \rangle t))/2.
	\end{equation}
	Now the amplitude scales as $1/k$ and the spin-dependent interaction determines the period $\pi/g_1\langle n \rangle$.
\end{description}
In the intermediate regime where both energies are of the same order of magnitude a spin dynamics resonance will occur which has been
observed for the first time in a harmonically trapped $^{87}$Rb $F=2$ spinor condensate in \cite{Kronjaeger2006a}.\\
\footnote{$F=2$ spinor condensates have an additional spin-dependent interaction parametrized by $g_2 \cdot |S_{0}|^{2}$.
However, due to the smallness of this contribution the resulting dynamics can well be approximated by the solutions found for $F=1$.}
Fig.\,\ref{fig:6} shows measurements for different values of $k$ for such a spin dynamics resonance.
Note that the tuning of the resonance has been achieved by changing the magnetic field value while keeping the atomic density as
constant as possible.

\subsection{Spin dynamics of $^{87}$Rb $F=1$ spinor condensates in a triangular lattice}

In this paper we discuss for the first time, that the mean-field spin dynamics resonance persists even in the presence of a periodic
potential as long as a significant superfluid fraction is still present in the system.
Moreover the renormalization of the spin-dependent interaction by the lattice potential allows for an effective tuning of the resonance
at constant magnetic field.\\
The experimental procedure for the results presented here is as follows.
After preparation of BEC in $|1,-1 \rangle$ with particle numbers of $N\approx 10^5$ and no discernible thermal fraction,
we load the atoms in the three-dimensional optical lattice consisting of a 2D triangular lattice and a perpendicular 1D standing wave as discussed above.
For these measurements the individual lattice depths have been adapted to yield equal $U/zJ$ for the different spatial directions.
After the final lattice depth has been reached, we allow the system to relax to equilibrium for $5\,\mathrm{ms}$ before
we apply a $\pi/2$-pulse to prepare the initial state $\mathbf{\zeta}_{\pi/2}$.
The magnetic field is then abruptly switched to the desired final value.
We then hold the atoms for a variable evolution time to allow spin dynamics to take place.
\begin{figure}[htb]
  \centering
  \includegraphics[width=1\textwidth]{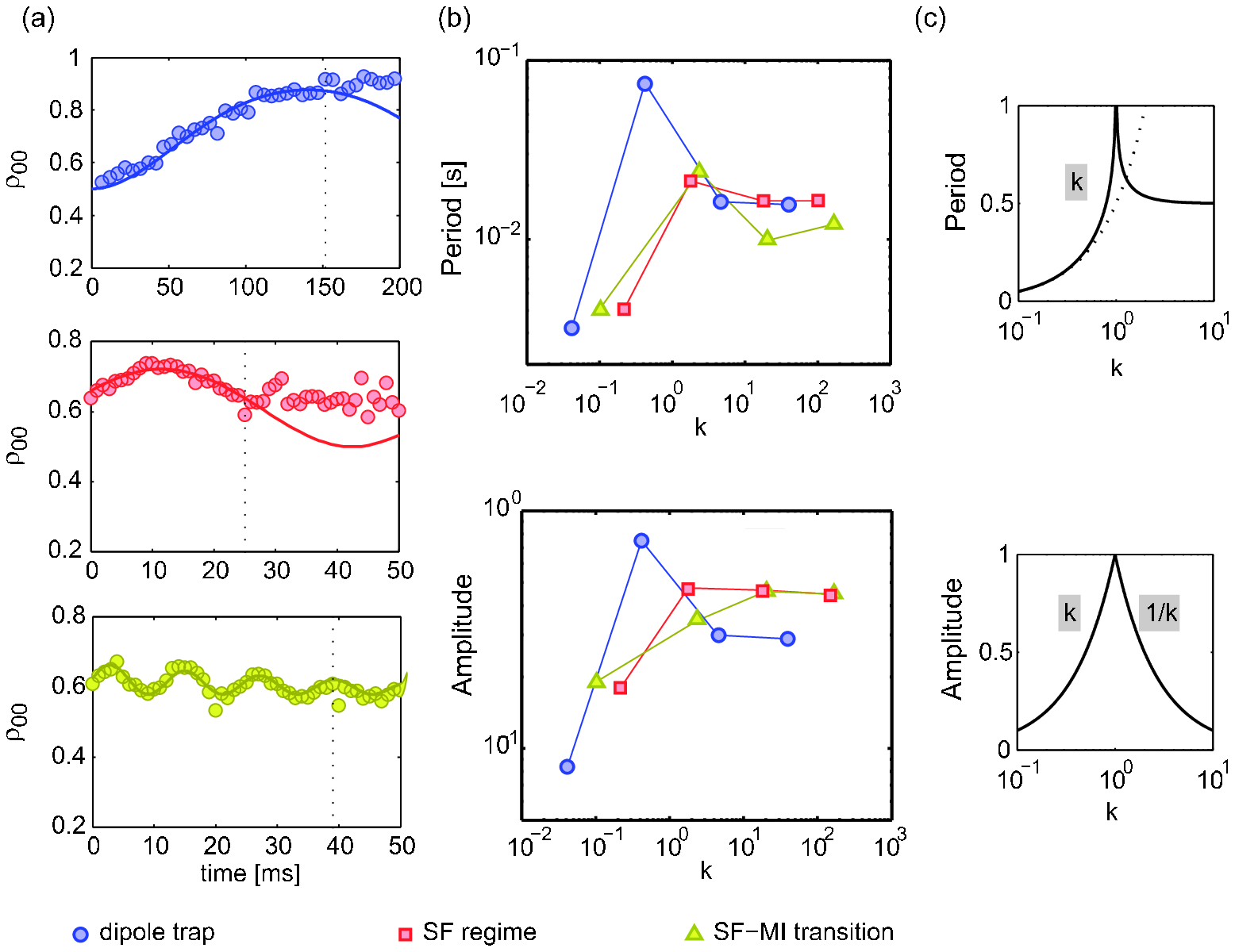}\\
  \caption[Figure7]{
  (a) Exemplary measurements of spin mixing dynamics for different strength of the periodic potential and different
  		magnetic fields.
  		From top to bottom measurements have been performed in the dipole trap at $|B|=270\,\mathrm{mG}$,
  		in the superfluid regime at $|B|=30\,\mathrm{mG}$ and close to the phase transition from superfluid to Mott-insulator
  		at $|B|=800\,\mathrm{mG}$.
  		The individual graphs show the time evolution of the population in the $m_{F} = 0$ state $\rho_{00}$.
			Data is presented along with fits (solid lines) to determine oscillation amplitude and period.
  (b) Period of spin mixing dynamics as a function of the dimensionless parameter $k_{\mathrm{eff}} = g_1\langle n_{\mathrm{eff}} \rangle / q$
  		for experiments in a crossed dipole trap (blue circles), in the superfluid regime in the presence of a triangular optical lattice (red triangles)
  		and close to the quantum phase transition from superfluid to Mott-insulator (yellow stars).
  		Solid lines are guide to the eye.
  (c) Theoretically expected period and amplitude according to Equ.\,\ref{equ:spinor:pop}.
  }
  \label{fig:7}
\end{figure}
In order to have atomic samples which are \emph{not} optically thick after time-of-flight, we quickly ramp down the lattice
within $5\,\mathrm{ms}$ following the end of the evolution time to obtain atomic clouds that allow for a reliable and unambiguous
atom number determination.
Finally all trapping potentials are switched off and after $21\,\mathrm{ms}$ of time-of-flight including Stern-Gerlach separation
the atoms are imaged with resonant light.\\
In the superfluid regime the addition of a periodic potential should mainly show up
in terms of an increased effective spin-dependent coupling owing to an increased mean density of the sample.
We expect that the resulting effective $k$ will determine frequency and amplitude of spin dynamics within this model.
However different additional effects e.g. caused by reduced mobility or an increased thermalization rate might lead to
deviations from this behavior and the data has been carefully analyzed if any indications for those deviations can be found.\\
To estimate atomic densities in the different lattice regimes we have employed a Thomas-Fermi-like approximation of the
Bose-Hubbard model for the superfluid regime and the regime close to the SF-MI phase transition.
The corresponding densities are calculated by taking into account the particular overall harmonic trapping frequencies $\bar{\omega}$,
onsite interaction matrix elements $U$ and total particle numbers $N$ according to $\bar{n} =4/7\,n_{0} = \mu/g_{0}$ with
\begin{equation}
\mu = \left( \frac{15}{16}\frac{V_{\mathrm{cell}} m^{3/2} N U \bar{\omega}^3} {\sqrt{2} \pi} \right)^{2/5},
\label{equ:meanDensity}
\end{equation}
Having obtained the mean density in this way a specific effective  $k$ value can be assigned to each set of measurements.
It should be mentioned, that our experimental parameters allow for a tuning of the effective density by a factor of approximately 5 by changing the lattice depth.\\
Fig. \ref{fig:7}a shows spin dynamics measurements in a crossed dipole trap (Hubbard parameter $\eta = 0$), in the superfluid regime ($\eta = 1.45$)
and in the vicinity of the SF-MI phase transition ($\eta = 8.9$) for exemplary magnetic fields.
It can be inferred from the graphs that the population in the $m_{F}=0$ component $\rho_{00}$ undergoes periodic oscillations.
The dynamics of the other components follows according to $\rho_{-1-1}=\rho_{11}=1/2\,(1-\rho_{00})$.
Oscillation amplitude and period clearly change depending on the lattice depth and the magnetic field.
The time evolution of the $m_{F} = 0$ population has been fitted with a function $\rho_{00} = (1 - A\sin^2(B \cdot (t + C)))/2$
to extract the oscillation amplitude $A$ and period $2\pi / B$.\\
A direct comparison of amplitudes and periods obtained in this way to the analytical mean-field solution Equ.\ref{equ:spinor:pop} shows indeed good agreement
if the spin-dependent interaction energy entering $k = g_1\langle n \rangle / q$ is renormalized by the effective density in the optical lattice as explained above (Fig. \ref{fig:7}b).
A major consequence of this observation is that by raising an optical lattice potential, the atomic density and thereby the effective $k$ can be varied which can readily be used
to precisely tune the spin dynamics resonance at constant magnetic field and constant particle number.
Beyond that we want to emphasize that the physics of spinor condensates in optical lattices demands for further investigation.
For example first indication for spin demixing at very low magnetic fields as well as its suppression with increasing lattice depth has been observed
in our experiments.
This effect leads to a slow-down and potentially stop of spin mixing dynamics.
The accurate and detailed analysis of this and similar effects is however beyond the scope of this paper and wil be presented elsewhere.

\section{Conclusion and Outlook}

In conclusion we have presented first experiments with ultracold atoms loaded in a triangular optical lattice.
In particular the MI transition has been investigated in a two dimensional system.
In direct comparison to a cubic lattice the transition occurs at much lower lattice depth due to the much stronger localization of the atoms.
Moreover we have observed spin dynamics of $^{87}$Rb $F=1$ spinor condensates in an optical lattice and found qualitative
agreement with the mean-field solution raised in \cite{Kronjaeger2005a} as long as the lattice is shallow enough to maintain a significant superfluid fraction.
The differences in the results at different lattice depths can mainly be reduced to an effective
spin-dependent interaction $g_{1} \langle n_{\mathrm{norm}} \rangle$ which allows for a tuning of the spin dynamics resonance.\\
Future perspectives include loading and investigation of spinor-BEC in the hexagonal anti-ferromagnetic lattice and analysis
of magnetic ground states.
Fascinatingly the sign of the tunneling matrix element $J$ can be reversed by slightly modulating the frequencies of the three laser beams creating the triangular
lattice as described in \cite{Eckard2009a}.
Thereby access to investigate fascinating model systems known from solid state
physics that give rise to effects like Neel ordering and spin liquid phases is possible.
Experiments concerning transport and spin-dependent tunneling for the creation of highly entangled multi-particle states
are highly interesting research directions.

\section{Acknowledgements}

We thank the DFG for funding within the Forschergruppe FOR801 and Graduiertenkolleg GrK1355.

\bibliography{paper}
\end{document}